# Correlation between morphology and transport properties of quasi-free-standing monolayer graphene

Yuya Murata,[1,*] Torge Mashoff,[2] Makoto Takamura,[3] Shinichi Tanabe,[3] Hiroki Hibino,[3] Fabio Beltram,[1,2] and Stefan Heun[1,†]

[1]*NEST, Istituto Nanoscienze-CNR and Scuola Normale Superiore, Piazza San Silvestro 12, 56127 Pisa, Italy*

[2]*Center for Nanotechnology Innovation @ NEST, Istituto Italiano di Tecnologia, Piazza San Silvestro 12, 56127 Pisa, Italy*

[3]*NTT Basic Research Laboratories, 3-1 Morinosato Wakamiya, Atsugi, Kanagawa 243-0198, Japan*

**Abstract**

We investigate the morphology of quasi-free-standing monolayer graphene (QFMLG) formed at several temperatures by hydrogen intercalation and discuss its relationship with transport properties. Features corresponding to incomplete hydrogen intercalation at the graphene-substrate interface are observed by scanning tunneling microscopy on QFMLG formed at 600 and 800°C. They contribute to carrier scattering as charged impurities. Voids in the SiC substrate and wrinkling of graphene appear at 1000°C, and they decrease the carrier mobility significantly.

[*] e-mail: yuya.murata@nano.cnr.it
[†] e-mail: stefan.heun@nano.cnr.it

Quasi-free-standing monolayer graphene (QFMLG) is obtained by intercalating H atoms at the interface between a buffer layer and SiC(0001).[1] The buffer layer consists of a honeycomb structure of C atoms formed by thermal decomposition of the SiC(0001) surface, covalently bonded to the substrate.[2] The carrier mobility of QFMLG shows less temperature dependence than graphene obtained by thermal decomposition of SiC(0001) (epitaxial monolayer graphene, EMLG).[3] This is attributed to a reduced interaction between QFMLG and the substrate. However, the mobility of QFMLG (~3000 $cm^2V^{-1}s^{-1}$) is limited to a value lower than exfoliated graphene on $SiO_2$. In order to identify strategies that can lead to improved QFMLG mobility it is necessary to better understand the carrier scattering mechanisms for this material. Some of us recently reported that the mobility of QFMLG depends on $T_H$, the substrate temperature during H intercalation.[4] The highest mobility was reported when $T_H \sim 700°C$. Furthermore, from the relationship between electrical conductivity and charge-carrier density, it was suggested that carrier scattering in these samples is mainly caused by charged impurities at $T_H = 600$ and $800°C$, while defects induce additional scattering at 950°C. Unfortunately the detailed nature of these defects could not be deduced from the transport data. In order to address this issue, here we report on the morphology of QFMLG formed at several $T_H$ values and discuss the relationship with their transport properties. Samples were imaged by scanning tunneling microscopy (STM), atomic force microscopy (AFM), and transmission electron microscopy (TEM).

The sample preparation process is identical to the one reported in Ref. 4, except for the sample discussed in Fig. 1(i). In brief, 4H- and 6H-SiC(0001) substrates were used as starting point. Samples were cleaned by annealing in $H_2$ at 33 mbar and 1500°C for 5 min. A buffer layer was formed by annealing in Ar at 800 mbar and 1650°C for 5 min. Finally, samples were annealed in $H_2$ at 1013 mbar and 600-1200°C for 60 min for H intercalation. STM was performed in ultra-high vacuum at room temperature using a RHK Technology STM with a base pressure of $1 \times 10^{-10}$ mbar. Images were obtained in constant current mode, i.e. a feedback adjusts the height of the STM tip above the sample surface such that the tunnel current is maintained constant. Samples were furthermore characterized by AFM in air at room temperature and by cross-sectional TEM.

Figure 1 shows STM results on samples obtained at $T_H$ = 600, 800, and 1000°C. A 6-fold honeycomb lattice indicating the formation of QFMLG is seen at all $T_H$ in the high-resolution images (Figs. 1(c), (f), and (i)). Bright spots and small dark spots are seen in samples obtained for $T_H$ = 600 and 800°C (see arrows in Figs. 1(a) and (b)). Fig. 1(b) shows both types of spots, and it is evident that the bright spots appear slightly larger than the dark spots. The apparent width (height) of the bright spots is a few nm (50 pm). A graphene $\sqrt{3} \times \sqrt{3}$ periodicity is resolved in the high-resolution STM image around a bright spot in Fig. 1(l) and its two-dimensional Fourier transform (inset). This closely resembles the $\sqrt{3} \times \sqrt{3}$ periodicity obtained in the simulated STM image of a defect in EMLG as a result of the characteristic inter-valley electron scattering.[5] We obtained similar STM images around defects on EMLG created by nitrogen ion sputtering.[6] We therefore identify the bright spots as defects in QFMLG.

The width and height of the small dark spots in these images are 1.5 nm and 15-25 pm, respectively. The appearance of them is constant in STM images with sample bias values ranging from +1.5 to -1.0 V. Within these spots a defect-free honeycomb structure is seen (see Figs. 1(c) and (f)). The small dark spots partially align along the SiC $\langle 11\bar{2}0 \rangle$ directions (see, for example, Fig. 1(b)) with a periodicity of 1.8 nm, as seen in Fig. 1(j), which is a cross-section along the red line shown in Fig. 1(b). This resembles a SiC(0001) quasi-(6×6) periodicity as seen in STM images of buffer layers before H intercalation[2] and suggests that small areas of incomplete H intercalation at the graphene-substrate interface are imaged as small dark spots. Contrast in the STM images could be for two reasons: either due to an electronic effect, or due to height variation of the surface. For an unscreened charged impurity (like a dangling bond) one would expect a contrast inversion upon bias variation. This is not observed, as stated above. This indicates a strong screening of the interface charge by the graphene film. We therefore conclude that the observed contrast is mainly due to height variation of the surface.

The height difference between QFMLG and a region where H atoms were desorbed by vacuum annealing (which reversibly returns the QFMLG to the buffer layer)[1] is approximately 100 pm.[7] On the other hand, the corrugation of the buffer layer is approximately 60 pm.[2] The depth of the small dark spots is smaller than both of these values, compared in STM images with similar sample bias voltages. We deduce that the dark spots do not correspond to buffer layer inclusions in the QFMLG. We argue that at the position of a small dark spot, a Si atom without H termination is isolated and surrounded by other H-terminated Si atoms. It does not covalently bond to graphene owing to steric hindrance. Such unsaturated Si atoms with a dangling bond align in a SiC(0001) quasi-(6×6) periodicity, as do the original covalent bonds of the buffer layer.

For the samples prepared at $T_H = 1000°C$, in addition to the features discussed so far, large dark spots appear. The width and height of these spots are 4-10 nm and 0.25 nm, respectively. This is shown in Fig. 1(k), which is a cross-section along the red line in Fig. 1(h). Their height corresponds to the height of a single SiC(0001) layer. The large dark spots are distributed randomly. Figure 1(i) shows a magnified image around a large dark spot on a sample which was annealed at 950°C in $H_2$ at 113 mbar.[7] The honeycomb lattice of QFMLG is observed also within the large dark spot (left side of the image). It continuously covers the edge of the large dark spot. From these observations, we conclude that the large dark spots are voids in the SiC substrate below the graphene that have the thickness of one SiC(0001) layer. The SiC may be etched during H intercalation at high temperature. The existence of QFMLG indicates that H intercalation took place also within these areas.

Graphene wrinkling is visible in the AFM images shown in Fig. 2(a-d).[8,9,10] This is particularly visible with samples produced at higher $T_H$. Wrinkles appear above $T_H = 800°C$, but they are not frequently seen on samples at $T_H = 800°C$ and 950°C. They are reproducibly formed on samples above $T_H = 1100°C$. Figure 2(e) shows a cross-sectional TEM image of one such wrinkle: wrinkling of two layers is visible. This is a quasi-free-standing bilayer graphene formed by H intercalation to an EMLG region which was partially grown together with a buffer layer. The width and height of wrinkles are approximately 3 nm and 1.8 nm, respectively. They may stem from the difference in the thermal expansion coefficients of graphene and SiC.[11]

We can now discuss the correlation between the observed QFMLG morphology and their transport properties. In photoelectron spectroscopy measurements,[12] a reduction of hole doping and a broadening of spectra was observed following QFMLG annealing in vacuum. This was interpreted as a consequence of partial H desorption at the graphene-substrate interface and formation of individual Si dangling bonds, leading to the formation of graphene in the free-standing state. In this scenario, these Si dangling bonds can donate charge to the graphene film and act as charged scattering centers. This is consistent with our interpretation that the small dark spots observed in the STM images are Si dangling bonds at the interface produced by incomplete H intercalation: they are the charged impurities which reportedly cause carrier scattering.[4]

The large dark spots on samples at $T_H = 1000°C$ are similar to a structure where graphene goes over a substrate step. In fact we argued that these spots correspond to voids in the SiC substrate. Scanning tunneling potentiometry on EMLG showed that substrate steps can increase the resistivity of graphene substantially compared to graphene on a perfect terrace.[13] The origin of the step-induced resistance may be a σ-π hybridization due to the curvature of graphene, or strain in graphene, and/or a reduced doping from the substrate due to a larger distance between graphene and substrate.[14] Wrinkles of graphene formed at high temperature also induce such curvature and strain in graphene, and a larger distance between graphene and substrate. Consequently they can be expected to have a similar influence on transport in QFMLG.

Figure 3 shows the density of bright, as well as small and large dark spots estimated from STM images on samples obtained at $T_H$ = 600, 800, and 1000°C. The different data points at a given temperature show the variation of the density of the features in STM images taken at different positions of a sample. The density of small dark spots at $T_H = 1000°C$ is two orders of magnitude smaller than for $T_H$ = 600 and 800°C samples. As temperature increases, the dissociation of $H_2$ molecules, the intercalation of H atoms, and their diffusion along the graphene-substrate interface may be promoted, and more H atoms are intercalated into the interface. At $T_H = 1000°C$ large dark spots and wrinkles are observed rather than small dark spots. Since the mobility at $T_H = 1000°C$ is lower than that at 600 and 800°C,[4] we attribute this lowering to SiC voids and/or graphene wrinkles. Two opposing trends contribute to the change in carrier scattering. At lower $T_H$ intercalation is incomplete and this leads to carrier scattering driven by charged impurities. Figure 3 shows that indeed this can be reduced by increasing $T_H$. On the other hand, higher $T_H$ leads to the formation of etch pits in the substrate and wrinkles of graphene. These have the opposite effect, i.e. a reduction in mobility. At the sample preparation conditions in this experiment, the formation of SiC voids and graphene wrinkles have a larger influence on the mobility offsetting the impact of increased H intercalation brought by increasing $T_H$.

Assuming that each bright spot is a defect in graphene, the density of defects is $5 \times 10^{11}$ cm$^{-2}$, i.e. less than 0.02%. This is one order of magnitude smaller than the calculated density of defects that opens a band gap in QFMLG, 0.3%.[15] Furthermore, the density of the bright spots is nearly constant between $T_H$ = 600°C and 1000°C. Importantly, defects in graphene are not the main reason for the observed decrease of the mobility at 1000°C.

In summary, we found that Si dangling bonds due to incomplete H intercalation at the graphene-substrate interface cause carrier scattering as charged impurities in QFMLG at $T_H$ = 600 and 800°C. At $T_H$ = 1000°C, pits in the SiC substrate and wrinkles of graphene appear and decrease the mobility of QFMLG, despite a better H intercalation. We conclude that a higher mobility of QFMLG can be obtained by optimizing the conditions for H intercalation while staying below the temperature at which pits and wrinkles appear.

We acknowledge financial support from the CNR in the framework of the agreement on scientific collaboration between CNR and JSPS (Japan), joint project title 'High-mobility graphene monolayers for novel quantum devices', and from the Italian Ministry of Foreign Affairs, Direzione Generale per la Promozione del Sistema Paese. We also acknowledge funding from the European Union Seventh Framework Programme under grant agreement n°604391 Graphene Flagship.

**Figure captions**

FIG. 1. (Color online) STM images on quasi-free-standing monolayer graphene. H atoms were intercalated at a temperature $T_H$ of (a-c) 600°C, (d-f) 800°C, and (g-i and l) 1000°C. Black arrows in (a) and (b) indicate a bright and a small dark spot, respectively. Blue arrow in (e) indicates a SiC(0001) <11$\bar{2}$0> direction. Scan size, bias voltage, and tunneling current are (a) 200 nm, -0.6 V, 0.5 nA, (b) 50 nm, 0.5 V, 0.5 nA, (c) 8 nm, -0.6 V, 0.4 nA, (d) 200 nm, 0.6 V, 0.5 nA, (e) 50 nm, 0.6 V, 0.4 nA (f) 8 nm, 0.5 V, 0.1 nA, (g) 200nm, 0.6 V, 0.5 nA, (h) 50 nm, 0.8 V, 0.4 nA, (i) 8 nm, -0.1 V, 0.6 nA, (l) 8 nm, 0.8V, 0.4 nA. Inset in (l) shows a two-dimensional Fourier transform of (l). White and blue arrows indicate a graphene $1\times1$ and $\sqrt{3}\times\sqrt{3}$ spot, respectively. (j) Line profile along red line in (b). (k) Line profile along red line in (h).

FIG. 2. (Color online) (a-d) AFM images of quasi-free-standing monolayer graphene. H atoms were intercalated at $T_H$ = (a) 650°C, (b) 800°C, (c) 950 °C, and (d) 1100°C. Scan size 14 μm. Images were differentiated in a horizontal direction for better contrast. (e) Cross-sectional TEM image of a sample obtained at $T_H$ = 1200°C.

FIG. 3. (Color online) Density of bright spots as well as small and large dark spots as a function of temperature for H intercalation.

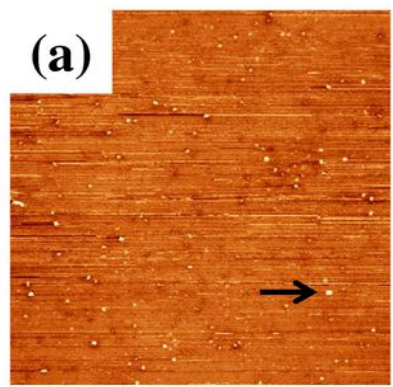 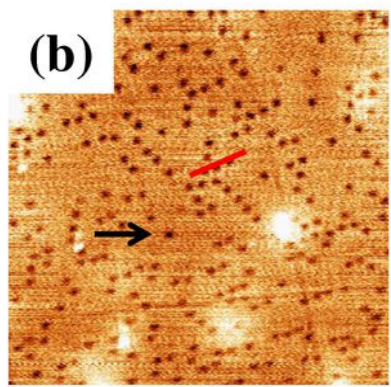 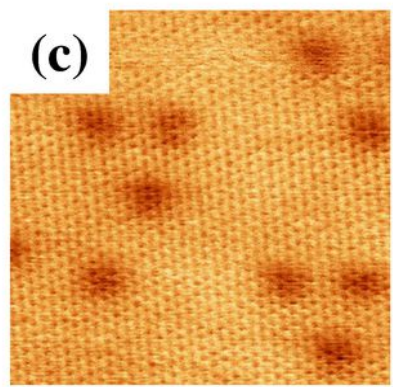 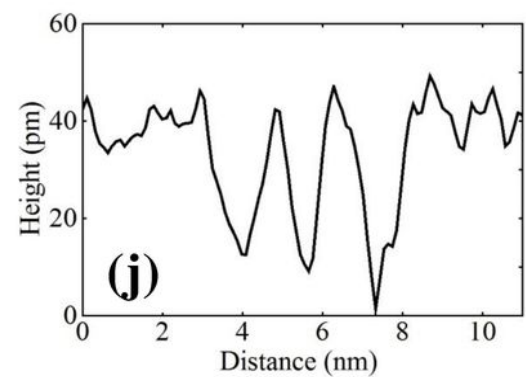
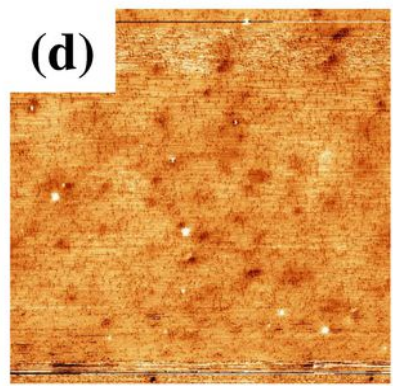 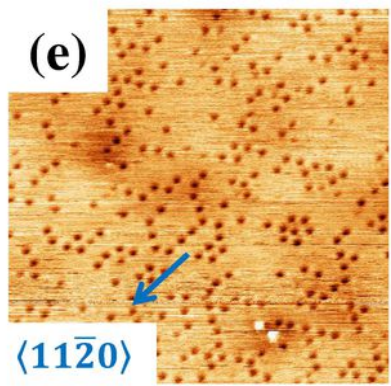 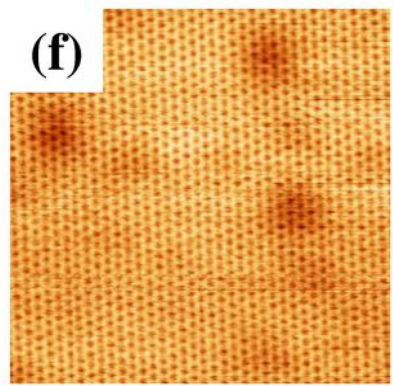 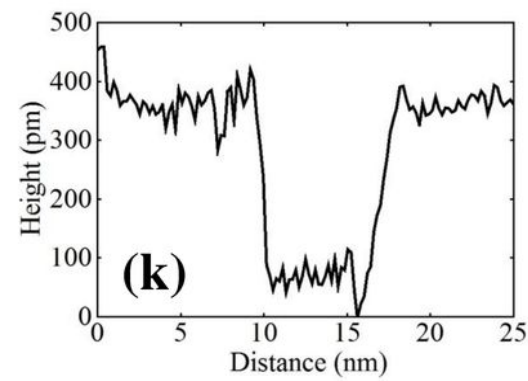
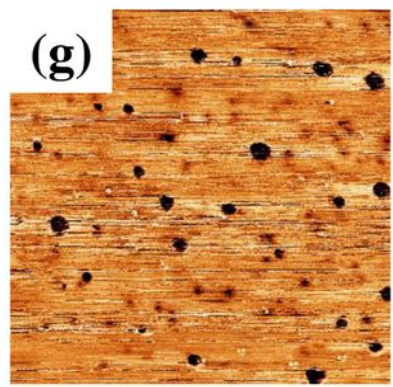 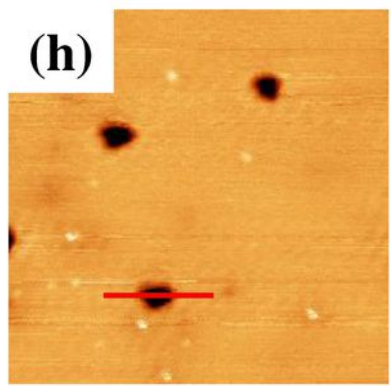 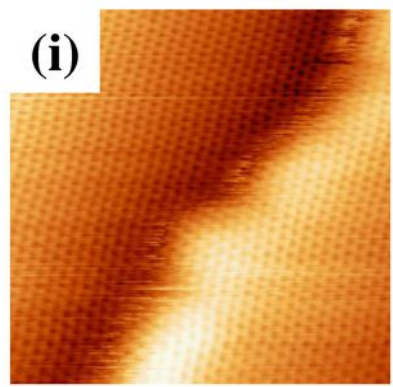 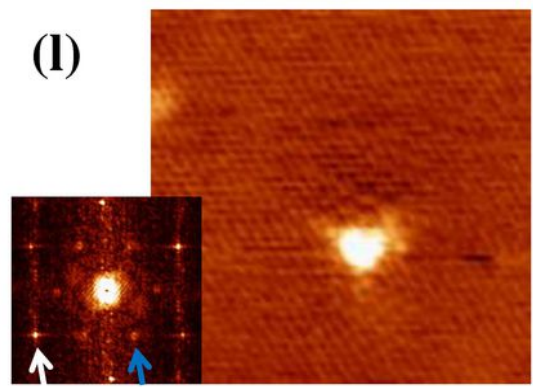

⟨11$\bar{2}$0⟩

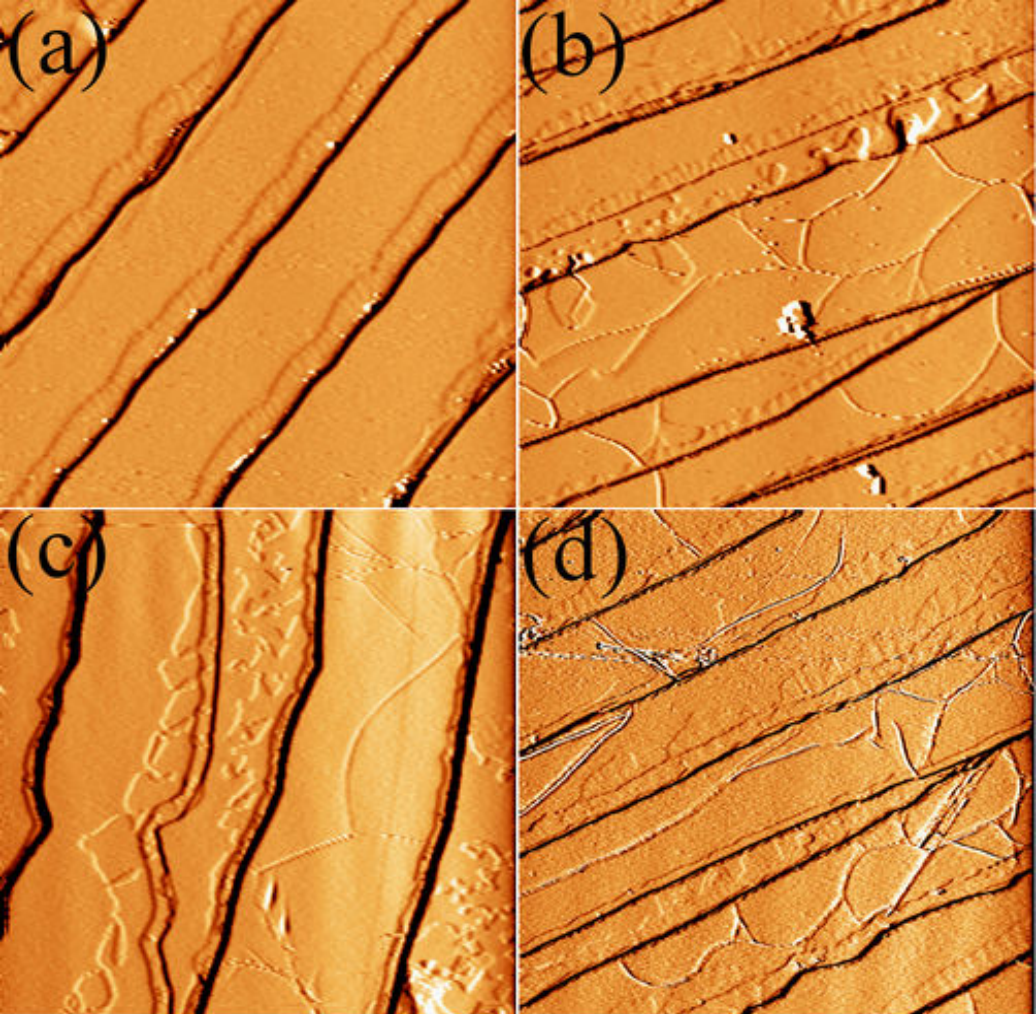

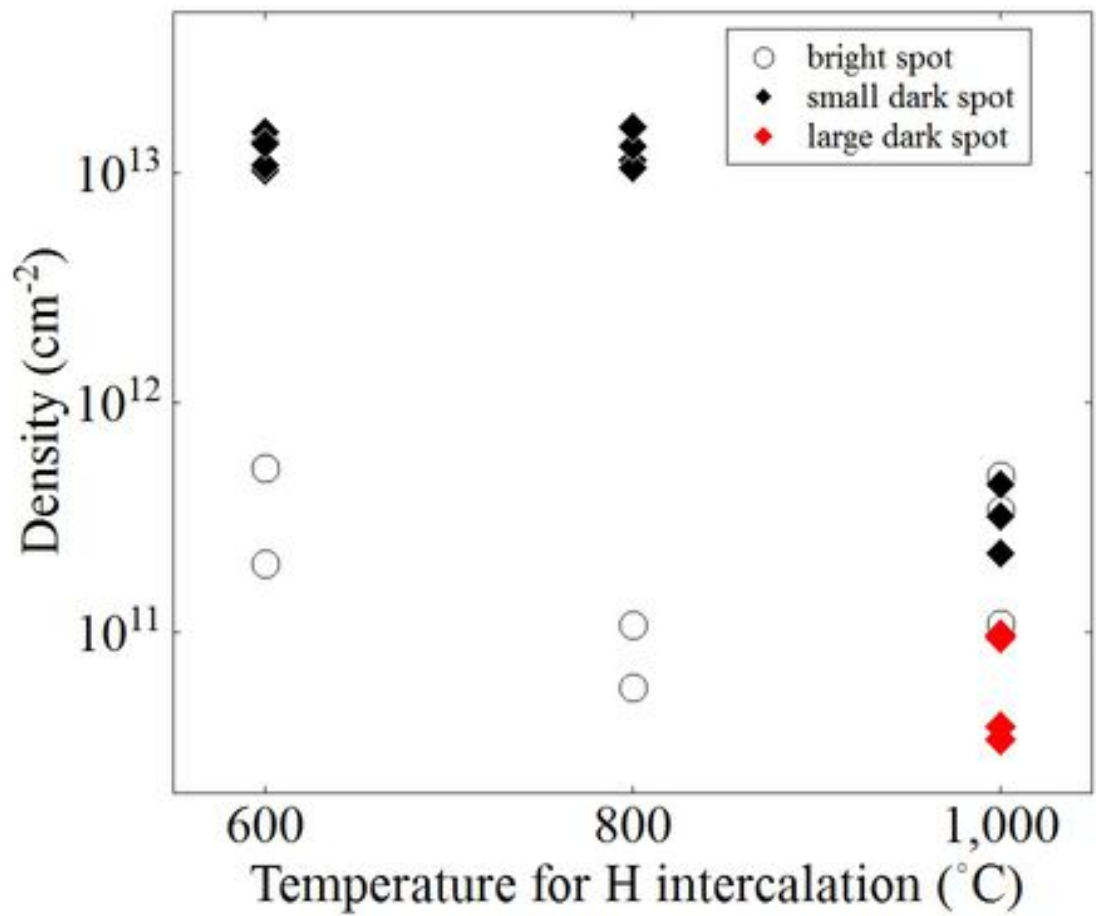